\documentclass[manuscript]{aastex}

\slugcomment{Not to appear in Nonlearned J., 45.}

\shorttitle{Thermal Instability behind a Shock Wave in HI and Molecular Clouds}
\shortauthors{Aota, Inoue $\&$ Aikawa}

\begin{document}


\title{Thermal Instability behind a Shock Wave in HI and Molecular Clouds}


\author{Takuhiro Aota\altaffilmark{1}, Tsuyoshi Inoue\altaffilmark{2}, Yuri Aikawa\altaffilmark{1} }

\altaffiltext{1}{Department of Earth and Planetary Sciences, Kobe University, Kobe 657-8501,Japan}
\altaffiltext{2}{Department of Physics and Mathematics, Aoyama Gakuin University, Fuchinobe, Chuou-ku, Sagamihara 252-5258, Japan}







\begin{abstract}
We performed one-dimensional hydrodynamic simulations with detailed cooling, heating and chemical processes
to examine the thermal stability of shocked gas in  cold neutral medium (CNM) and molecular clouds.
We find that both CNM and molecular clouds can be thermally unstable in the cooling layer behind the shock wave.
The characteristic wavelength of the thermal instability ranges from $10^{-5}$ pc to $0.1$ pc in the  CNM,
and from $10^{-7}$ pc to $0.1$ pc in the molecular clouds. This coincides with the size of observed tiny scale structures
in the  CNM and molecular clouds, indicating that the thermal instability in the shocked gas  could be a formation mechanism of these tiny structures in the interstellar medium.
We have also calculated the $e$-folding number of the thermal instability to estimate the amplification of the density fluctuation in the shocked gas.
Density perturbations in the  CNM grow by a factor of exp(5)$ \simeq 150$,
whereas the perturbations in the molecular clouds grow only by a factor of a few behind a high Mach number shock. The amplification factor is larger at lower densities
and higher velocities.
Formation of very small-scale structures by thermal instability in shocked gas is more effective in lower densities.

\end{abstract}


\keywords{ISM: clouds -- ISM: molecules -- ISM: structure -- Shock waves }



\section{Introduction}
Thermal instability is an important physical process to determine the structure in the interstellar medium \citep{f65}.
It is well established that the neutral gas in the interstellar medium (ISM) consists of two distinct phases:
cold neutral medium (CNM) and warm neutral medium (WNM) (e.g.,  \citealt{fgh69}; \citealt{whmtb95, ht03}).
\cite{fgh69} calculated the thermal equilibrium state in ISM
considering the cosmic ray heating and line cooling by H, O, and C[II].
They showed that there are three physical states under the pressure equilibrium: two stable states and one unstable state.
The stable states correspond to the CNM and the WNM.
When the WNM transforms to the CNM in converging flows and/or shocks, the gas goes through a thermally unstable state (e.g., \citealt{hp99, ki00};
 \citealt{hbhsd05, vrpgg06}).

Many authors have studied the dynamical condensation and fragmentation processes of the ISM driven by the thermal instability in the shock in WNM.
\cite{ki00, ki02} showed that small-scale clumps of CNM are formed behind the shock front in WNM (see also, \citealt{hp99, ha07, ii08, ii09, hhsdb08, vgj07}).
\cite{ki00} also investigated the shock propagation {\it within}  the CNM, and found that the shocked layer in  the CNM is thermally unstable, as well.
More recently, \citet{ii12} has succeeded in forming very turbulent molecular clouds by accretion of  CNM mixed with WNM.

The thermal instability works as follows. Consider an isobaric gas with small density perturbation.
If the gas in small density enhancement (i.e. temperature decline) has a larger cooling rate than the surrounding gas, the density enhancement grows.
In other words, the condition is determined by how the cooling rate depends on the density and temperature.
Thermal instability thus could occur in denser regions like molecular clouds  as well, and has actually been studied.
\cite{ddb80} investigated the thermal-chemical instability (e.g., \citealt{gl76}) in the chemical transition region where the dominant form of carbon
changes from C[II] to CO (\citealt{gl75}), and showed that the thermal-chemical instability is not active.
\cite{g84} investigated thermal instability in molecular clouds with density $n \sim 10^3$ cm$^{-3}$ 
and temperature $T \simeq 35$ - $75~{\rm K}$, and found that such a cloud is thermally unstable.
\cite{n07, n11} investigated thermal instability in molecular cloud cores ($n \sim 10^5 - 10^6$ cm$^{-3}$) with the effect of the ambipolar diffusion,
and showed that the thermal instability can grow in quasi-magnetohydrostatic, self-gravitating slab (\citealt{n07}) and axisymmetric cylindrical core (\citealt{n11}).

On the other hand, to our best knowledge, the role of thermal instability in shock heated molecular clouds has not yet been studied.
Molecular clouds are characterized by supersonic velocity dispersions, which are most probably due to turbulence
(\citealt{l81, srby87}).
Dissipation of the supersonic turbulence would be accompanied by shocks. Shock waves are also driven by collisions of protostellar outflows with ambient gas.

Molecular clouds are known to have clumpy structures.
While many of the clumps are gravitationally bound, there are very small-scale structures as well: a size of $\sim$1000AU and density of $n \simeq 10^{4}$ cm$^{-3}$ \citep{lvk95, h02, ss03, tea12}.
Since they are gravitationally unbound ($M \leq 0.05 M_{\odot}$), they cannot be formed by gravitational instability.
Although shock compression by turbulence often makes gravitationally unbound structure,
it would not be easy to make very small-scale structure ($<0.1$ pc) by shock compression alone.
Firstly, the shock compresses the gas as a whole, and thus cannot make fragments.
Since turbulence has eddies of various spatial scales, one may imagine that the shock compression at small scale eddied can make very small-scale structures or fragments.
But it should be noted that the turbulence (i.e. velocity dispersion in molecular clouds) becomes subsonic at $<$ 0.1 pc according to Larson's law (see e.g., \citealt{hb04}).
Structures smaller than this scale thus cannot be formed by shock compression by turbulence.
If the shocked molecular gas is thermally unstable, it could generate very small-scale fragments.
The effect of thermal instability on the shocked molecular gas must be explored.

In this paper, we examine the thermal stability of shocked gas in molecular clouds using one-dimensional hydrodynamic simulations including detailed cooling, heating and chemical processes.
Although our main target is molecular clouds, we also calculate one model of  CNM, in order to compare our results
with \citet{ki00}, and to compare the e-folding numbers in  CNM and molecular clouds.
This paper is organized as follows.
In \S 2, we describe our physical and chemical models.
Then, we explain the condition for the thermal instability and how we evaluate the growth of perturbation in \S 3.
In \S 4, we show the results of our simulations of shock propagation in  CNM and molecular clouds.
Finally, we summarize our results in \S 5.

\section{The model}
We investigate the evolution of ISM swept by a shock wave in a plane-parallel  gas.
Figure \ref{cal-picture} schematically shows the configuration of our model;
we consider a collision of oppositely oriented gas flows which have the same density, temperature and
chemical composition.
The external radiation irradiates both ends of the  numerical domain.
We calculate the temporal variation of temperature, density, and chemical composition in the shocked region.

\subsection{Basic Equations}
We solve the following equations: 
\[ \partial_{t}U(t,x)+\partial_{x}F_{x}=S  \]
\[ U=(\rho,\rho v_{x},E) \]
\[ F_{x}= \left( \begin{array}{ccc} \rho v_{x} \\ 
\rho v_{x}^{2} + p \\  
 (E+p) v_{x} \end{array} \right) \]
\[ S= \left( \begin{array}{ccc} 0 \\
0 \\
\rho\,(\Gamma-\Lambda) \end{array} \right) \]
\[ E=\frac{p}{\gamma - 1} + \frac{\rho v^{2}_{x} }{2} \quad ,\]
where $\rho, v_{x}, p$, $\gamma$, $ \Gamma $ and $\Lambda$ are the gas mass density, velocity, thermal pressure, ratio of specific heat,
heating and cooling rate per unit mass, respectively.
We use an operator-splitting technique to solve these equations, which are split into three parts: (1) ideal hydrodynamics,
(2) cooling and heating, and (3) chemical reactions (e.g., \citealt{ii08}).
The first part, ideal hydrodynamics, is calculated by employing a second-order Godunov method with Lagrangian coordinates \citep{v79}.
We solve the exact Riemann problem iteratively at each grid cell interface to calculate numerical fluxes,
and determine the position of grid cell interface in the next time step.
Thus, we can appropriately calculate small-scale compressed dense regions and large-scale pre-shock regions at once.
The energy equation
\[ \frac{\partial E}{\partial t} =  \rho\,(\Gamma-\Lambda)  \]
is solved by the second-order explicit method.
Temporal variation of the number density of chemical species is determined by the rate equations
\[ \frac{d n_i}{d t} = \sum_{j} k_{ij} n_{j}+\sum_{j, l} k_{ijl} n_{j} n_{l} , \]
where $k$ is the rate coefficient of the chemical reactions. The rate equations are calculated by a first-order implicit method \citep{h09}.

The time step of the integration is set to be small enough to satisfy the CFL condition, 
and to be much smaller ($\lesssim 4$ \%) than the cooling time scale of the gas.
The details of the physical and chemical processes considered in our model are described in the following.

\subsection{Heating and Cooling Processes}
A full list of the thermal processes included in our model is listed in Table \ref{tab:cool-heat}.
Our model includes the cooling by the line emission of H, C[II], O[I], and CO with the effect of radiative trapping, recombination on grains, and dust-gas collision.
We adopt the formula of escape probability by \cite{ddb80} for the line cooling by C[II] and O[I], and by \cite{hm79} for CO.
The escape probability is a function of the column density of the molecule or atom $N_{i}=\int\,n_{\rm i}dx$;
the number density is integrated from the edge of the  numerical domain to the center of each grid cell.
Since our model is 1D, we calculate the escape probabilities towards the right-hand-side and left-hand-side edges of the  numerical domain, 
and use the average of these two values. Gas is also cooled  by collisions with dust;
the dust temperature is set to 10K (e.g., section 5 of \citealt{t05}).
Heating processes include photo-electric heating by PAH \citep{bt94}, cosmic ray \citep{gl78}, and H$_{2}$ photo-dissociation \citep{bd77}.
In the calculation of photo-electric heating, we consider the attenuation of external radiation.
Visual extinction, $A_{\rm V}$, is calculated by
\[ A_{\rm V} = \frac{ \int{n_{\rm H}dx} }{ 1.89 \times 10^{21} ~ {\rm cm^{-2}}} ~~{\rm mag}  \]
where  the numerator is the column density of hydrogen nuclei integrated from the edge of the numerical domain to the center of each grid cell (e.g., \citealt{mmp83}).

\subsection{Chemical reactions}
We calculate the chemical reaction network in the gas phase, which consists of 462 species and 9578 reactions. 
Chemical reactions and rate coefficients are adopted mainly from OSU network
(http://www.physics.ohio-state.edu/\verb|~|eric),  which is developed for interstellar chemistry.
Namely, we use the network of \cite{gh06} at $T\lesssim 100$K
and \cite{hhw10} (see also \citealt{hhw12}, the errata of \citealt{hhw10}) at $T>100$K. 
These networks contain cosmic ray ionization, ion-molecule reactions, neutral-neutral reactions, recombination of ion,
photo reactions, and grain surface reactions.
We also include the collisional dissociations in Table A1 of \cite{wkmh98},
but some rate coefficients are modified (see the appendix of \citealt{fat12}).
We do not consider grain-surface reactions except for H$_{2}$ formation.
Cosmic-ray ionization rate is set to be 1.3 $\times 10^{-17} $ s$^{-1}$.
In the model of  CNM gas, the total column density of  hydrogen nuclei is only $N_{\rm H}=1.0 \times 10^{20}$ cm$^{-2}$,
and the molecules are destroyed by photo reactions.
We take into account the self-shielding effects of H$_{2}$, CO and C atom referring to \cite{lhprl96} and \cite{th85};
the shielding factors are given as a function of $A_{\rm v}$ and the column densities of specific species integrated from
the edges of the  numerical domain to each grid cell at every time step.
Elemental abundances in the gas phase are listed in Table \ref{tab:elemental abundance}.

\subsection{Initial and Boundary conditions}
The initial temperature and chemical composition of the colliding gases are determined by calculating the thermal 
and chemical equilibrium.
In the model of  CNM, these initial conditions vary spatially depending on the visual extinction $A_{\rm v}$ at each position in the  numerical domain.
In the model of molecular clouds, $A_{\rm v}$ is set to be 5 mag for all  grid cells;
our numerical domain is an embedded small portion of the molecular cloud.
In total, we calculate one  CNM model and 45 molecular cloud models.
The model parameters are summarized in Table \ref{tab:initial state}.

\begin{figure}
\epsscale{.60}
\begin{center}
\fbox{
\plotone{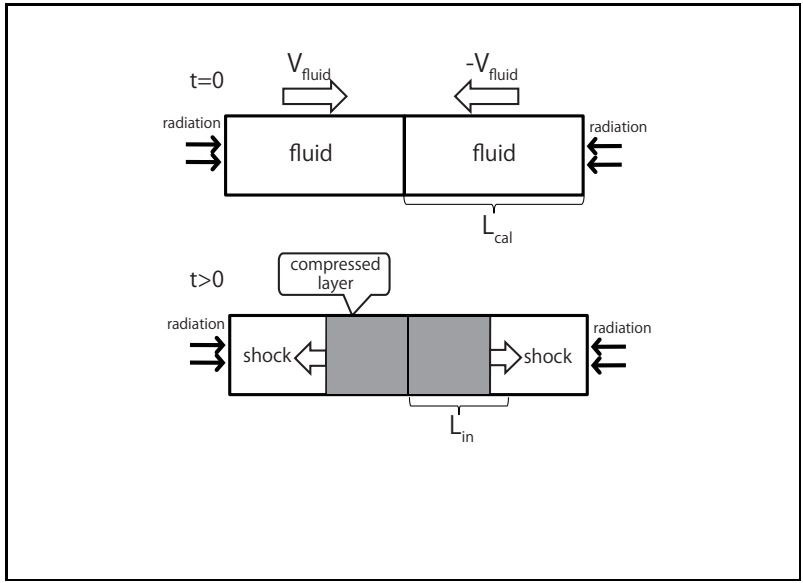}
}
\end{center}
\caption{Schematic view of our 1D shock model}
\label{cal-picture}
\end{figure}

Our initial condition of the  CNM is basically the same as that of \citet{ki00} (see their \S 3.4);
we assume the number density to be $n_{\rm H}=10$ cm$^{-3}$.
Thermal equilibrium determines the gas temperature, which is about 110 K at any position.
The velocity of the colliding gas $V_{{\rm fluid}}$ (see Figure \ref{cal-picture}) is 10km/s,
which corresponds to the velocity of gas flow associated with old supernova remnants.
Total column density of hydrogen nuclei is set to be $1.0 \times 10^{20}$ cm$^{-2}$, i.e. $A_{\rm v} \simeq 0.05$ mag.

For the molecular cloud models, we explore the parameter space of the number density $n_{\rm H}$:
$1.0\times 10^2$, $3.0 \times 10^2$, $1.0\times 10^3$, $3.0 \times 10^3$ and $1.0\times 10^4$ cm$^{-3}$.
The velocity $V_{{\rm fluid}}$ ranges from 0.5 to 4.5 km s$^{-1}$, referring to the turbulent velocities
in molecular clouds \citep{l81, fkm08}.
We also investigate a model with $V_{\rm fluid}=10$ km s$^{-1}$, which is a typical velocity of protostellar outflows
(e.g., \citealt{ms88}).

In the model of the  CNM, the size of our numerical domain $L_{\rm cal}$ is 3.24 pc, which covers the whole region of 
the  CNM. In order to calculate the post-shock region with a high spatial resolution, we set 225  grid cells at
$x < L_{\rm in}$, and 35  grid cells at $x > L_{\rm in}$ (see Figure \ref{cal-picture}). Initially, $L_{\rm in}$ is
0.972 pc. Both $L_{\rm cal}$ and $L_{\rm in}$ change with time, since we are using Lagrangian coordinates.
In the models of molecular clouds, our numerical domain varies from 0.01 pc to 0.28 pc depending on the model parameters
(see Table \ref{tab:initial state}). The numerical domain is divided into
300  grid cells with equal intervals.

We adopt the free-boundary condition;
\[ \frac{\partial \rho}{\partial x}=0,~~~ \frac{\partial p}{\partial x}=0,~~~ \frac{\partial v_{x}}{\partial x}=0, \]
i.e. we assume that the physical values of left and right side of boundary are the same when we solve the exact Riemann problem at boundary.
Basically, the boundary condition is not important, because we stop the calculation before the shock wave reaches the boundary.

\section{Amplitude of thermal instability}
\cite{b86} showed that for any unperturbed state, the gas is thermally unstable if
\begin{equation}
\left[ \frac{\partial}{\partial s} \left( \frac{\Lambda - \Gamma}{T} \right) \right]_{A} < 0,
\label{balbus}
\end{equation}
where $s$ is specific entropy, $T$ is temperature, and $A$ is a thermodynamic variable kept constant in the perturbation.
Since the ISM is mostly in pressure equilibrium, the isobaric condition, $A=p$, is satisfied, which leads to
\begin{equation}
\left[ \frac{\partial}{\partial T} \left( \frac{\Lambda - \Gamma}{T} \right) \right]_{p} < 0.
\end{equation}

\cite{sms72} and \cite{ki00} performed a linear analysis of thermal instability in isochorically cooling gas and 
isobarically contracting gas, respectively.
When the gas is thermally unstable, the density perturbation grows as $\rho=\rho_0\, \exp(\sigma \,t)$.
The growth rate in isobarically contracting gas is
\begin{eqnarray}
    \sigma &=&-\frac{m\,T\,(\gamma-1)}{\gamma\,k_{\rm B}} \left[ \frac{\partial}{\partial T} \left( \frac{\Lambda - \Gamma}{T} \right) \right]_{p}\nonumber\\
    &=& \frac{1}{\gamma} \left\{ \frac{ 1 + s_{\rho} - s_{T} }{ \tau_{ {\rm cool} } } - \frac{ 1 + r_{\rho} - r_{T} }{ \tau_{ {\rm heat} } } \right\}, \label{dispersion}
\end{eqnarray}
\[ s_{\rho}=\partial ( {\rm ln}~\Lambda ) / \partial ( {\rm ln}~n ) ,\quad s_{T}=\partial ( {\rm ln}~ \Lambda) / \partial ( {\rm ln}~T ), \]
\[ r_{\rho}=\partial ({\rm ln}~ \Gamma) / \partial ({\rm ln}~n) , \quad r_{T}=\partial ({\rm ln}~ \Gamma) / \partial ({\rm ln}~n), \]
where $m$ is the mean molecular mass, and $\tau_{ {\rm cool}}\equiv k_{\rm B}\,T/(\gamma-1)/(m\,\Lambda)$ and
$\tau_{ {\rm heat} }\equiv k_{\rm B}\,T/(\gamma-1)/(m\,\Gamma)$ are the cooling and heating time scales, respectively.
 $\sigma$ is a function of density and temperature, that changes with time as the gas goes through the shock wave and enters the cooling region.
When the gas is thermally unstable, $\sigma$ takes a positive value.

In this paper, we evaluate the integrated $e$-folding number $\int \sigma dt$ as an indicator of the amplification of the perturbation,
where the $e$-folding number is calculated for each fluid element in Lagrangian coordinate,
and the integral is executed only when the Balbus criterion (Eq \ref{balbus}) is satisfied.
Then, using the $e$-folding number, we obtain the amplification of the thermal instability $\exp(\int \sigma dt)$.

In the linear analysis, the growth rate is a function of the wavelength of the perturbation,
and Eq. (\ref{dispersion}) is the rate for the most unstable mode at
\begin{equation}
\lambda_{\rm max}=\sqrt{l_{{\rm F}} l_{{\rm a}}}.
\end{equation}
The Field length $l_{\rm F}$ is 
\begin{equation}
l_{\rm F}= \left\{ \frac{\kappa\,T}{\rho\,(\Gamma-\Lambda)} \right\}^{1/2},
\label{field}
\end{equation}
where the thermal conductivity $\kappa$ is 2.5$\times 10^{3} T^{0.5}$ cm$^{-1}$ K$^{-1}$ s$^{-1}$ \citep{p53}.
The acoustic length $l_{\rm a}$ is
\begin{equation}
l_{\rm a}=c_{\rm s}\left( \frac{e}{\Gamma - \Lambda}\right) ,
\label{acous}
\end{equation}
where $c_{\rm s}$ is the sound speed, $e$ is the specific internal energy, and $ e/(\Gamma - \Lambda) $ is the net cooling time scale \citep{f65}.
It should be noted, however, that the dependence of the growth rate on perturbation wavelength is considerably weak around the most unstable wavelength.
The growth rate is comparable to equation (\ref{dispersion})
in the wavelength range of
\begin{equation}
l_{\rm F} < \lambda < l_{\rm a}
\end{equation}
(\citealt{f65}, see also Appendix B of \citealt{ki00}).

Ideally, the growth of density perturbation should be measured in hydrodynamic simulations starting from an
initial condition with small amplitude perturbations. But it is not easy in practice. 
First of all, the thickness of the unstable layer is comparable to or smaller than the most unstable wavelength
(see Figure 3).
In one dimensional flow, the perturbation grows as long as Eq. (1) is satisfied, but when the gas is compressed to be in thermally stable state, the perturbation is dispersed, which is an artifact.
In 2D and/or 3D simulations, large scale perturbations up to the acoustic length $l_{\rm a}$ can grow in directions
{\it parallel} to the shock front, and these clump structures remain in the post-shock regions  (e.g., \citealt{ii08}).
The 2D/3D simulation is, however, very time consuming, especially if we are to resolve the perturbations with very small wavelengths.
It is well established that the growth rate derived from the linear analysis is applicable to non-linear regime
of thermal instability.
 For instance, we can see in Figure 1 of \cite{iik07} that the isobaric condition, which is required for the thermal instability to grow with the linear growth rate, is met throughout the evolution
without interceptive feedback effects.
Therefore, it is reasonable to estimate the amplification of the density perturbation by integrating the {\it e}-folding number along the 1D flow.
It should at least be done before investing a large computational time on 2D hydrodynamic simulation with perturbation and very high spatial resolution.
We also note here that our present work is analogous to \cite{ki00}; they performed 1D shock calculation to find that the gas becomes thermally unstable in the shock-compressed layer
and predicted that small clumps would be formed. 
Later, \cite{ki02} indeed showed that such clumps are formed in the 2D simulation.

\section{Results}
We have calculated the generation and propagation of shock waves in  CNM and molecular clouds by solving the
basic equations in \S 2.1. In the following, we show the spatial distribution of physical parameters and
the {\it e}-folding number when the shock wave reaches a steady state.

\subsection{Shock Propagation in  CNM}
\begin{figure}
\epsscale{0.80}
\plotone{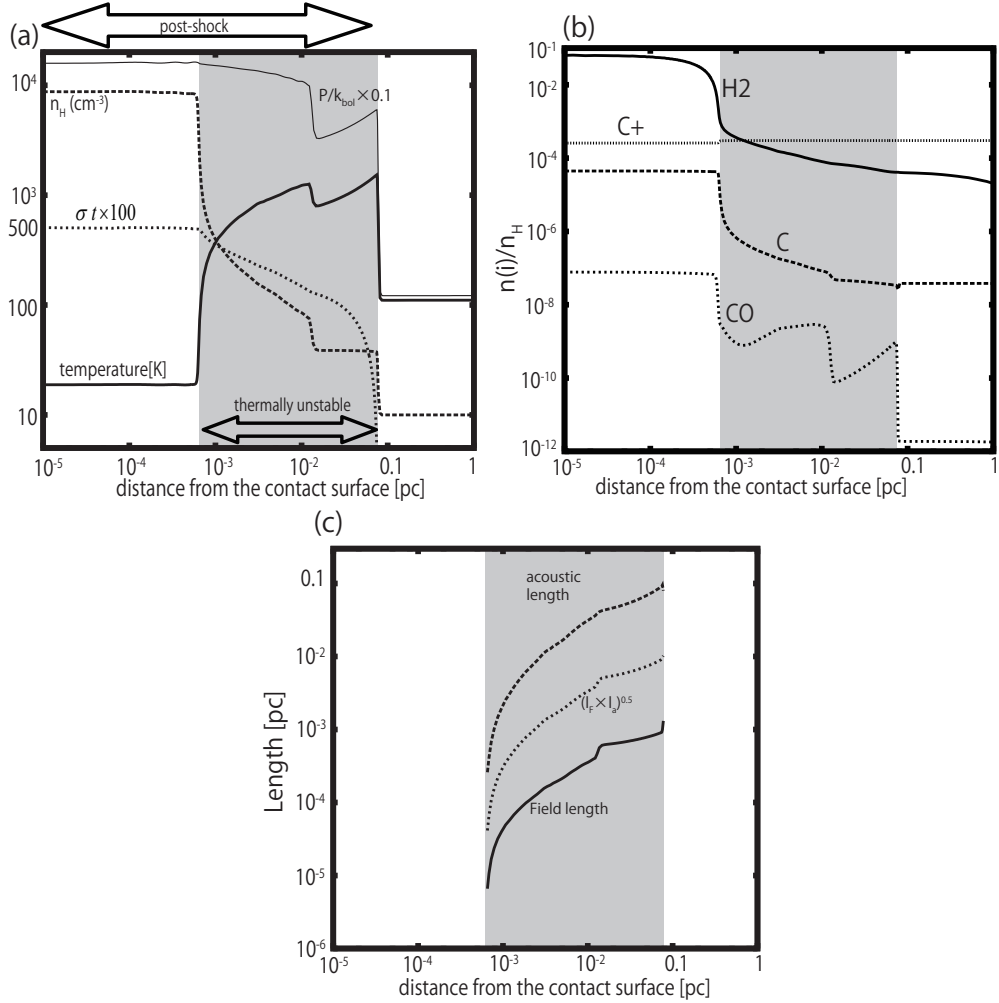}
\caption{(a) Spatial distribution of temperature ({\it solid line}), number density of hydrogen nuclei
({\it dashed line}),  thermal pressure ({\it thin-solid line}), and integrated $e$-folding number $\sigma t$ ({\it dotted line}) at $t=88000$ yr.
Thermally unstable region is shaded in gray.
(b) Spatial distribution of H$_{2}$, C, CO and C$^+$ abundances with respect to hydrogen nuclei at $t=88000$ yr.
(c) Spatial distribution of Field length, acoustic length and most unstable wavelength $\sqrt{l_{\rm F} l_{\rm a}}$.}
\label{diffuse}
\end{figure}

Figure \ref{diffuse} (a) shows the distribution of temperature, number density of hydrogen nuclei ($n_{{\rm H}}$),  thermal pressure,
and integrated $e$-folding number at $t=88000$ yr in the  CNM model.
 A similar simulation was performed by \cite{ki00}.
Temperature and density distributions in Figure 2 are indeed similar to Figure 7 of \cite{ki00},
who found that the shocked  CNM evolves through a thermally unstable state.
The gray shade in Figure \ref{diffuse} depicts the thermally unstable region, in which the condition (\ref{balbus}) is satisfied.
CNM becomes thermally unstable immediately behind the shock and then evolves to a thermally stable dense gas
with basically isobaric condition (see Figure 7 (a) of \citealt{ki00}).
The main coolants are CII(158$\mu$m) and OI(63$\mu$m).
Now, we go one step further from \cite{ki00} and calculate the {\it e}-folding number.
The integrated $e$-folding number $\int \sigma dt$ is approximately 5, which means that the density fluctuation grows
by a factor of exp(5)$\simeq 150$.
Note that the density profile shown in Figure \ref{diffuse} is the unperturbed value.
For example, if the pre-shock gas has a density fluctuation of 10$\%$, i.e. 11 cm$^{-3}$ in the pre-shock gas of 10 cm$^{-3}$,
the fluctuation grows by a factor of 150, 1.6 $\times$ 10$^{4}$ cm$^{-3}$ in the post-shock gas of 1.0 $\times$ 10$^{3}$ cm$^{-3}$.

We also show the distribution of assorted chemical species (Figure \ref{diffuse} $b$).
Molecular hydrogen in the pre-shock region in our model is more abundant than that of \cite{ki00} due to the
difference in the self-shielding model; \cite{ki00} used the formulation by \cite{th85}, while
we use the Table 10 in \cite{lhprl96}. In the dense stable region ($x \lesssim 10^{-3}$ pc), on the other hand,
the H$_2$ abundance in our model is almost the same as that in \cite{ki00}.
H$_{2}$ is formed by the association of H atoms on grain surfaces and destroyed by photo-dissociation.
Figure \ref{diffuse} (b) also shows the abundances of C$^{+}$, C and CO.
Although \cite{ki00} did not show the spatial distribution of CO, they reported that 0.02 \% of the carbon is in CO
in the dense stable gas in the post-shock region.
In our model 0.03 \% of carbon is in CO. It should be noted that we solve the detailed chemical network, whereas
\cite{ki00} adopted a simplified chemical model that assumed a direct conversion of C$^+$ to CO without accounting explicitly for the intermediate reactions.
Consistency of our model results with \cite{ki00} validates their
simplified model.

Figure \ref{diffuse} (c) shows the Field length, acoustic length and the most unstable wavelength of 
perturbation $\sqrt{l_{{\rm F}} l_{{\rm a}}}$ in the unstable region.
The Field length is $10^{-5} - 10^{-3}$ pc and the acoustic length is $10^{-3} - 10^{-1}$ pc.
The most unstable wavelength ranges from $10^{-4}$ to $10^{-2}$ pc, which coincides with the size of the
tiny scale structures observed in  CNM, $\sim 10^{-3}$ pc (see Table 1 of \citealt{h97}).
Such tiny structures can be formed by thermal instability \citep{ki00}.

\subsection{Shock Propagation in Molecular Clouds}
\begin{figure}
\epsscale{.90}
\plotone{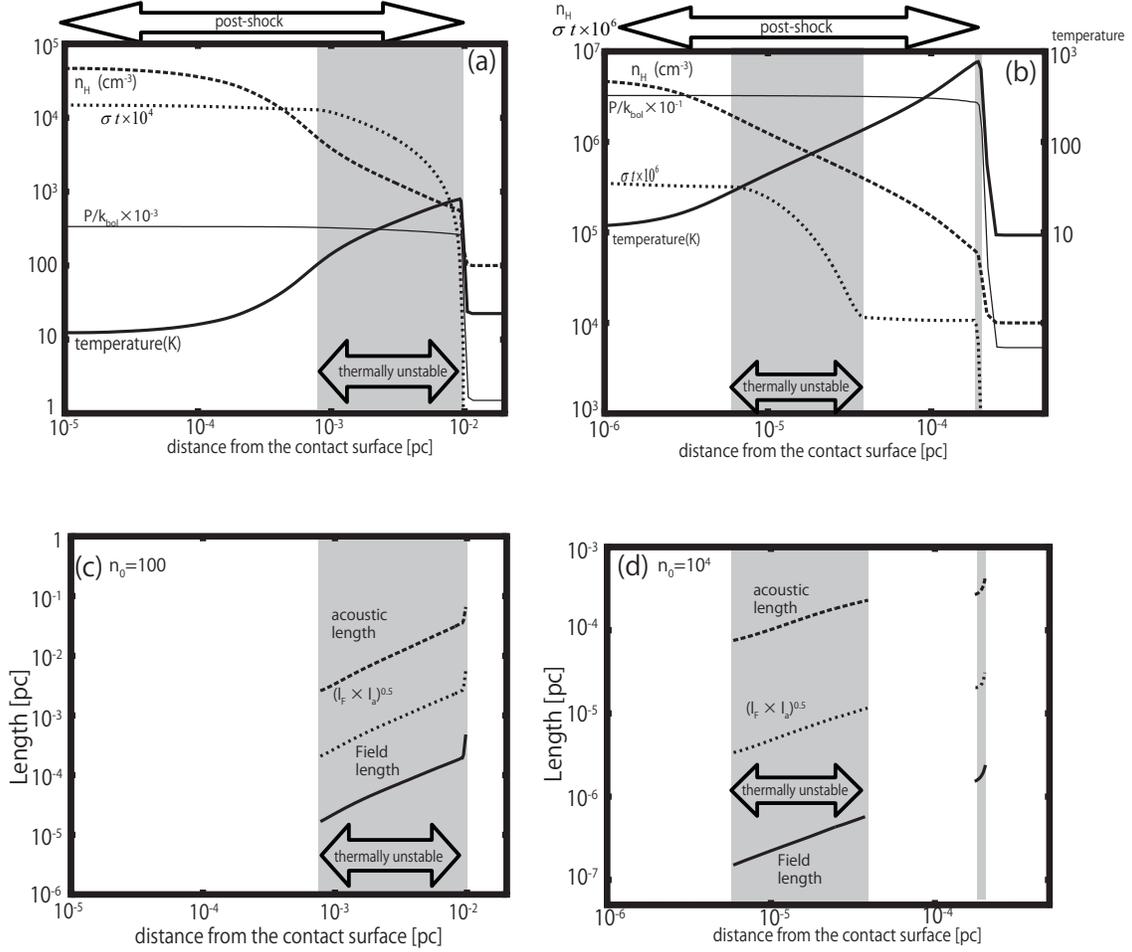}
\caption{Spatial distribution of temperature ({\it solid line}), number density of hydrogen nuclei ({\it dashed line}),
 thermal pressure ({\it thin-solid line}), and integrated $e$-folding number ({\it dotted line}) in the model with $n_{\rm H}=10^{2}$ cm$^{-3}$ at
$t=53000$ yr (a), and in the model with $n_{\rm H}=10^{4}$ cm$^{-3}$ at $t=1500$ yr (b).
(c), (d) Spatial distribution of Field length, acoustic length and the most unstable wavelength
$\sqrt{l_{\rm F} l_{\rm a}}$ in the models with high and low initial densities.
The thermally unstable region is shaded in gray, as in Figure 2.}
\label{molecular}
\end{figure}

Figure \ref{molecular} (a) shows the spatial distribution of temperature, number density of hydrogen nuclei ($n_{{\rm H}}$),
 thermal pressure, and integrated $e$-folding number
$\int \sigma dt$ in the model with 4.5 km s$^{-1}$ and a pre-shock density $n_{\rm H}=100$ cm$^{-3}$ at $t=53000$ yr.
The gas becomes thermally unstable immediately behind the shock, where the gas temperature  reaches near 1000 K.
 The gas evolves in the post-shock region with the isobaric condition.
The main coolant is CO in this warm gas. The integrated $e$-folding number is 1.25;
if we calculate the HD simulation with a small perturbation, it grows only by a factor of 3.5.

Figure \ref{molecular} (b), on the other hand, shows a model with a higher initial density 
$n_{\rm H}=1\times 10^4$ cm$^{-3}$ at $t=1500$ yr.
In this model, the post-shock gas becomes thermally unstable right behind the shock front, but then becomes stable,
although the temperature (several hundreds of K) in this region is similar to that in the unstable post-shock gas in Figure \ref{molecular} (a).
Once the number density $n_{\rm H}$ reaches several times 10$^{5}$ cm$^{-3}$,
the dust-gas collisional cooling becomes dominant and the gas becomes thermally unstable again.
The integrated $e$-folding number is even smaller than in the model of Figure \ref{molecular} (a).

Figure \ref{molecular} (c) and (d) show the spatial distribution of the Field length, acoustic length and the most unstable wavelength
$\sqrt{l_{\rm F} l_{\rm a}}$ in the two models of molecular gas.
In the model with higher gas density, the cooling rate is higher (see below), and thus these scale lengths
become shorter (see Eq (\ref{field}) and (\ref{acous})).

\begin{figure}
\begin{center}
\epsscale{.90}
\plotone{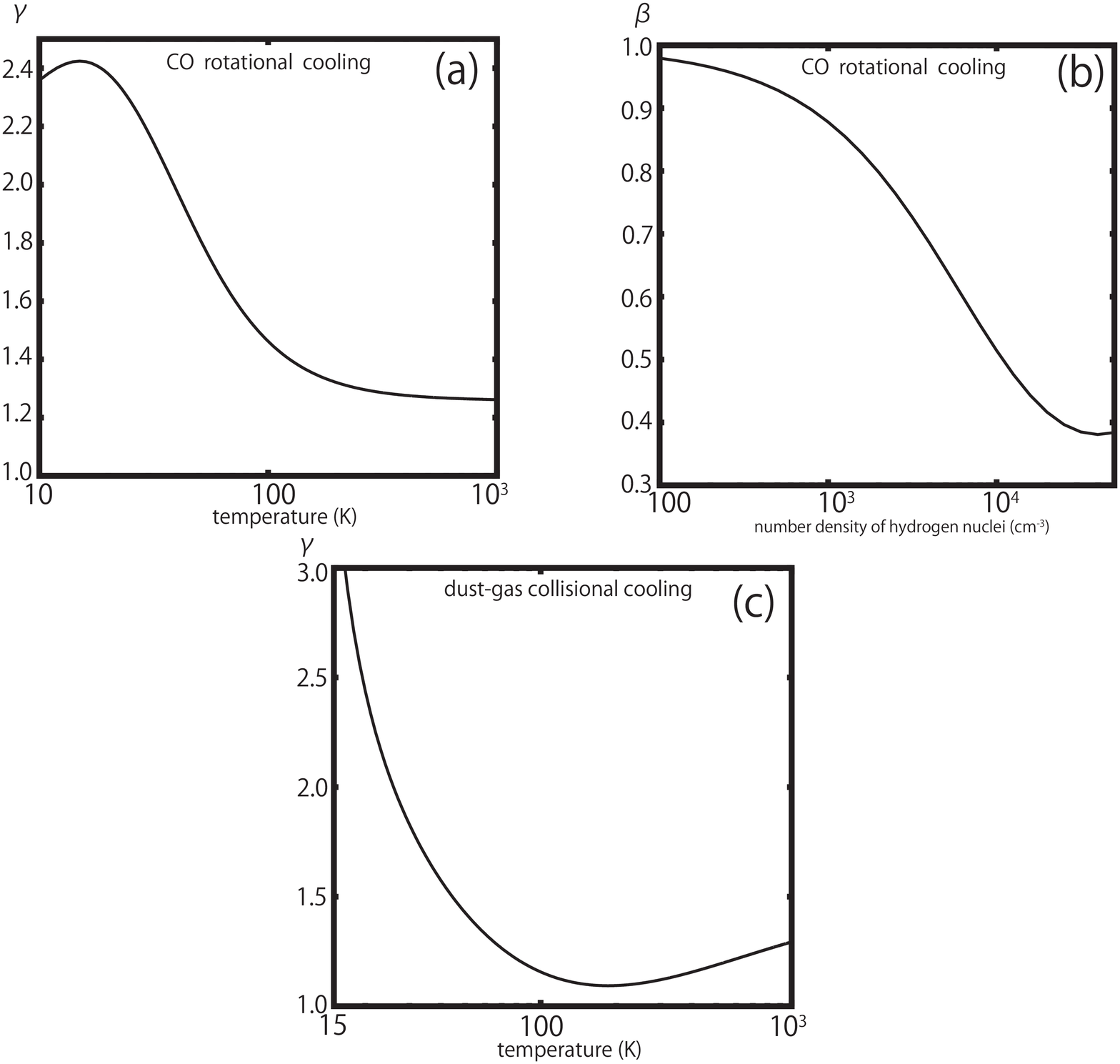}
\end{center}
\caption{(a)-(b) The power-law index of temperature, $\gamma= \partial {\rm ln}(\Lambda)/ \partial {\rm ln}(T)$, is calculated at a given density $n_{{\rm H}}=10^{3}$ cm$^{-3}$, CO abundance of $1.6\times 10^{-4}$ and CO column density  of $N_{{\rm CO}}=7.0\times 10^{17}$. The power-law index of density, $\beta=\partial{\rm ln}(\Lambda)/\partial{\rm ln}(n)$, is calculated at $T=100$ K, CO abundance of $1.6\times 10^{-4}$ and $N_{{\rm CO}}=7.0\times 10^{17}$.
(c) The power-law index $\gamma$ of the dust-gas collisional cooling is calculated at $T_{{\rm dust}}=10$ K.}
\label{CO-cooling}
\end{figure}

We summarize the integrated $e$-folding number in our molecular cloud models in Table \ref{tab:molecular-result}.
We can see that the shocked molecular gas is unstable when $V_{\rm fluid}$ is larger than $\sim 
1.5\,\log(n_{\rm H}\,[\mbox{cm}^{-3}])-2$ km s$^{-1}$, and that the integrated $e$-folding number increases with increasing $V_{\rm fluid}$
and decreasing initial gas density. These dependences can be understood as follows.
When the net cooling rate per unit mass $L$ is proportional to $n^{\beta}T^{\gamma}$, the criterion for thermal
instability by \cite{b86} can be rewritten as $\beta - \gamma +1 > 0$ (see Appendix A).
The gas is more unstable when $\beta$ is larger and $\gamma$ is smaller.
In other words, the perturbation grows faster when the dependence of the cooling rate on the temperature is weaker
and/or dependence of the cooling rate on gas density is stronger.
 The cooling rate actually is a more complicated function of density and temperature than a power law $n^{\beta} T^{\gamma}$.
But we can define $\beta$ and $\gamma$ as a local tangent in logarithmic plot.
Figures \ref{CO-cooling} (a) and (b) show the local tangent, $\beta$ and $\gamma$, of the cooling rate by CO  rotational lines
as a function of temperature and density at a typical molecular cloud condition.
We can see that $\beta$ is larger at lower densities and $\gamma$ is
larger at low temperatures.
In general, when the gas temperature is high enough to excite the coolant species (e.g., C$^{+}$, CO),
the cooling rate does not significantly decrease as the temperature is lowered.
When the gas temperature is comparable to the upper state energy of the line, on the other hand, the cooling rate decreases
steeply as the gas temperature is lowered.
The dependence of $\beta$ on density can be understood by considering the critical density, which is $3.3 \times 10^{6} (T/1000)^{0.75} $
cm$^{-3}$ for CO  rotational lines \citep{mswg82}.
When the gas density is much smaller than the critical density, the cooling rate per unit volume is proportional to $n^{2}$,
whereas at the critical density or higher, the dependence is weaker than $n^{2}$ because of collisional de-excitation 
(see e.g., section 2.3.1 of \citealt{t05}).

In Table \ref{tab:molecular-result}, asterisks indicate
that the dust-gas collisional cooling dominates over the CO cooling in the model. The cooling rate 
by dust-gas collision per unit mass is given as
\begin{eqnarray}
 1.2 \times 10^{31} n_{{\rm H}} \left( T_{{\rm gas}}/1000 \right)^{0.5} (100 {\rm \AA}/a_{{\rm min}}) \times \left[ 1- 0.8 {\rm exp}(-75/T_{{\rm gas}}) \right] (T_{{\rm gas}}-T_{{\rm dust}})/m \ \ {\rm erg} \ {\rm g}^{-1} \ {\rm s}^{-1},\nonumber
\end{eqnarray}
where $m$ is the mean molecular mass and we set $a_{\rm min} = 100$ {\AA} and $T_{{\rm dust}}=10$ K \citep{hm89}.
It is obvious that $\beta$ is always 1 in any density region,
and that dust-gas collisional cooling is important at high densities.
Figure \ref{CO-cooling} (c) shows the power index $\gamma$.
The cooling rate is proportional to $T_{{\rm gas}}^{1.5}$ when $T_{{\rm gas}} >> T_{{\rm dust}}$, but
$\gamma$ becomes larger than 1.5 when $T_{{\rm gas}} \simeq T_{{\rm dust}}$ (see Appendix B).
Since the dust temperature is 10 K in our model, the gas is more unstable at higher temperatures.

\section{Summary and Discussion}
We performed the one-dimensional hydrodynamic simulations with the effects of heating, cooling and chemical reactions
in order to study the thermal stability of shocked gas in  CNM and molecular clouds.
Taking advantage of the fact that the growth rate derived from the linear analysis \citep{sms72, ki00} is applicable to
the non-linear regime in thermal instability, we calculate the $e$-folding number along the flow to evaluate the
amplification of density perturbation behind the shock wave. Our findings are as follows
\begin{itemize}
\item{Both  CNM and molecular cloud can be thermally unstable behind a shock wave.}
\item{A molecular cloud becomes thermally unstable behind a shock when $V_{\rm fluid}\gtrsim 1.5\,\log(n\,[\mbox{cm}^{-3}])-2$ km s$^{-1}$.}
\item{The integrated $e$-folding number in the shocked molecular cloud increases with increasing $V_{\rm fluid}$ and decreasing pre-shock density.}
\item{The wavelength, $l_{\rm F}\lesssim \lambda \lesssim l_{\rm a}$, the perturbation of which can grow
within the cooling time scale, ranges from $10^{-5}$ pc to $0.1$ pc in the  CNM,
and from $10^{-7}$ pc to $0.1$ in molecular clouds. 
The unstable wavelength is a decreasing function of pre-shock density and fluid velocity, since both the Field length
$l_{\rm F}$ and acoustic length scale $l_{\rm a}$ decrease with gas density.}
\end{itemize}

The unstable wavelength of the thermal instability coincides with the size of the tiny scale structures observed in the  CNM \citep{h97, sh05} and molecular clouds \citep{lvk95, h02, ss03, tea12}.
Thermal instability {\bf could thus} explain the formation of such small gravitationally-unbound clumps in the ISM.
In the  CNM, the initial perturbation is amplified by a factor of $10^2$ in the thermally unstable region behind a shock.
In molecular clouds, on the other hand, the initial perturbation is amplified only by a factor of a few. 
It should be noted, however, that the super-sonic velocity dispersion is ubiquitous in molecular clouds.
Small clumps would be formed if the molecular cloud is swept by multiple shocks.

Finally, we discuss the fate of the structure formed by the thermal instability.
In the HI medium, the CNM can coexist with the WNM thanks to the thermally bistable nature (\citealt{fgh69}).
If the molecular clouds are isothermal uni-phase medium, the density fluctuations enhanced by the thermal instability could exist only in the very narrow shock transition layer, 
because the post-shock gas eventually returns to the same temperature as the pre-shock gas.
However, recent numerical simulations of molecular cloud formation (e.g., \citealt{bvhk09, ii12}) have shown that
molecular clouds are composed of the cold molecular gas ($T \sim 10$ K and $n >$ 100 cm$^{-3}$) and
the {\it non-equilibrium} diffuse warm gas ($T > 1000$ K and $n \sim$ 1 cm$^{-3}$).
The diffuse warm component, which is generated by the cloud-forming shocks at the envelope of molecular cloud, is in high pressure and thermally unstable.
The implication of this result is twofold. Firstly, it shows that a bistability of thermal equilibrium gas is not needed for density fluctuations to survive in post shock gas.
Secondly, in such a "non-equilibrium two-phase medium", the structure formed by the thermal instability behind the shock within molecular clouds would be more likely to
survive in the non-equilibrium diffuse gas.
 If the perturbed gas can fragment and coexist with the diffuse gas, the very small-scale structure could survive 
until at least the diffuse gas cools down ($\sim 0.1-1$ Myr).
But it is still an open question, and multi-dimensional simulations are necessary to confirm our expectation.

\acknowledgments
 We are grateful to the anonymous referee for helpful comments, which have improved the manuscript.
This work is supported by Grant-in-aids from the Ministry of Education, Culture, Sports, Science, and Technology (MEXT) of Japan, No. 21740146 and No. 23740154 (T. I.), and No. 21244021, No. 23540266, and
No. 23103004 (Y.A.). Numerical computations were in part carried out on Cray XT4 at Center for Computational Astrophysics, CfCA, of National Astronomical Observatory of Japan.

\appendix

\section{Modification of Balbus criterion}
We assume a gas with the cooling rate $A n^{\beta}T^{\gamma}$  (cf. \citealt{e91}).
The number density and temperature of an unperturbed state are $n_{0}$ and $T_{0}$.
Then the gas is compressed isobarically; the perturbed gas density and temperatures are $n_{p}=\alpha n_{0}$
($\alpha > 1$) and $T_{p}=T_{0}/\alpha$. Balbus criterion becomes
\begin{equation}
\frac{\partial}{\partial T} \left( \frac{\Lambda - \Gamma}{T} \right)_{p} \simeq \frac{ A T_{p}^{\gamma} n_{p}^{\beta}/T_{p} - A T_{0}^{\gamma} n_{0}^{\beta}/T_{0} }{T_{p} - T_{0}} <0 . \nonumber
\end{equation}
Considering $ T_{p} - T_{0} <0 $, $n_{p}=\alpha n_{0}$ and $T_{p}=T_{0}/\alpha$, it is straightforward to derive
\begin{equation}
A T_{p}^{\gamma} n_{p}^{\beta}/T_{p} - A T_{0}^{\gamma} n_{0}^{\beta}/T_{0} =
A T_{0}^{\gamma -1} n_{0}^{\beta} \left( \alpha^{\beta - \gamma + 1} - 1 \right)  > 0 . \nonumber
\end{equation}
We can see that the Balbus criterion is satisfied if $\beta - \gamma + 1 > 0$.

\section{Temperature dependence of dust-gas collisional cooling}
The cooling function of dust-gas collisional cooling is 
\begin{equation}
f[T]=AT^{0.5} \left( T - T_{{\rm dust}} \right),
\end{equation}
where $A$ is constant, and $T$ and $T_{{\rm dust}}$ are temperatures of gas and dust, respectively.
The power index $\gamma=\partial {\rm ln}(f[T])/\partial {\rm ln}(T)$ becomes
\begin{equation}
\gamma \simeq \frac{ {\rm ln}(f[(1+d\delta) T]) - {\rm ln}(f[T]) }{ {\rm ln}((1+d\delta) T) - {\rm ln}(T) },
\end{equation}
where $d\delta \simeq 0$. We set $T=\alpha T_{{\rm dust}}$, and it is straightforward to derive
\begin{equation}
\gamma \simeq 0.5 + \frac{ {\rm ln}(1+d\delta(1-1/\alpha)^{-1}) }{ {\rm ln}(1+d\delta) }.
\end{equation}
If $\alpha >>1 $($T >> T_{{\rm dust}}$), $\gamma \simeq 1.5$ and if $\alpha \simeq 1 $($T \simeq T_{{\rm dust}}$), $\gamma$ becomes larger than 1.5.




\clearpage



\begin{table}[htb]
\begin{center}
\small
\begin{tabular}{l l}

\hline
Cooling \& heating process & reference \\
\hline
Ly-$\alpha$ cooling & \cite{s78} \\
C$^{+}$ fine structure cooling(158$\mu$m) & \cite{ddb80}, \cite{hm89} \\
O fine structure cooling(63$\mu$m) & \cite{ddb80}, \cite{wmht03} \\
CO rotational cooling & \cite{hm79}, \cite{hi06} \\
CO vibrational cooling & \cite{hm89} \\
Cooling due to recombination on grains & \cite{bt94} \\
Cooling by the collision with dust & \cite{hm89} \\
Photoelectric heating by PAH & \cite{bt94}, \cite{wmht03} \\
Cosmic ray heating  & \cite{gl78} \\
H$_{2}$ photo-dissociation heating  & \cite{bd77} \\
\hline
\end{tabular}
\caption{Heating and cooling processes}
\label{tab:cool-heat}
\end{center}
\end{table}


\begin{table}[htb]
\begin{center}
\begin{tabular}{cc|cc}

\hline

element & abundance & element & abundance \\
\hline
He & 9.75$\times 10^{-2}$ & Fe & 2.47$\times 10^{-9}$ \\
O & 4.5$\times 10^{-4}$ & Na & 2.25$\times 10^{-9}$\\
C & 3.02$\times 10^{-4}$ & Mg  & 1.09$\times 10^{-8}$\\
N & 2.47$\times 10^{-5}$ & P  & 2.16$\times 10^{-10}$\\
S & 9.14$\times 10^{-8}$ & Cl  & 1.0$\times 10^{-9}$\\
Si & 2.47$\times 10^{-9}$ & &\\
\hline
\end{tabular}
\caption{Elemental abundance in the gas phase relative to hydrogen}
\label{tab:elemental abundance}
\end{center}
\end{table}



\begin{table}[htb]
\begin{center}
\begin{tabular}{ccccccccc}

\hline
{\it n$_{{\rm H}}$ } [cm$^{-3}$] & {\it T } [K] & G$_{0}$ \tablenotemark{a} & H atom \tablenotemark{b} & H$_{2}$ \tablenotemark{b} &
 C atom \tablenotemark{b} & CO \tablenotemark{b} & C$^{+}$ \tablenotemark{b} & $L_{{\rm cal}}$ \tablenotemark{c} [pc] \\
\hline
\multicolumn{9}{c}{CNM} \\
10 & 110 & 1.7 & 1.0 & 2.5(-5) & 3.8(-8) & 1.8(-12) & 3.0(-4) & 3.24 \\
\hline
\multicolumn{9}{c}{molecular cloud} \\
100 & 22 & 1.0 & 1.0(-2) & 0.49 & 1.2(-4) & 1.64(-4) & 1.5(-5) & 0.28 \\
300 & 15 & 1.0 & 3.0(-3) & 0.5 & 1.6(-6) & 3.0(-4) & 3.3(-7) & 0.15 \\
$1\times 10^{3}$ & 13 & 1.0 & 1.0(-3) & 0.5 & 3.8(-7) & 3.0(-4) & 1.1(-7) & 0.05 \\
3$\times 10^{3}$ & 10.5 & 1.0 & 4.2(-4) & 0.5 & 9.7(-8) & 3.0(-4) & 4.3(-8) & 0.02 \\
$1\times 10^{4}$ & 9.3 & 1.0 & 1.3(-4) & 0.5 & 2.3(-8) & 3.0(-4) & 1.4(-8) &  0.01 \\
\hline
\end{tabular}
\tablenotetext{a}{External FUV radiation normalized to the local interstellar radiation field (1.6 $\times 10^{-3}$ ergs cm$^{-2}$ s$^{-1}$) of \cite{h69}.}
\tablenotetext{b}{Relative abundance of chemical species to hydrogen nuclei($n_{{\rm H}}$). $a(-b)$ means $a\times 10^{-b}$.}
\tablenotetext{c}{Initial numerical domain (see Figure \ref{cal-picture}).}
\caption{Initial state}
\label{tab:initial state}
\end{center}
\end{table}


\begin{table}[htb]
\begin{center}
\begin{tabular}{|c|ccccc|}

\hline
V$_{{\rm fluid}}$& \multicolumn{5}{|c|}{$n_{{\rm H}}$ [cm$^{-3}$]} \\

[km s$^{-1}$]  & $10^{2}$ & $3 \times 10^{2}$ & $10^{3}$ & $3 \times 10^{3}$ & $10^{4}$  \\
\hline
0.5 & $\times$ & $\times$ & $\times$ & $\times$ & $\times$  \\
1.0 & 0.08 & $\times$ & $\times$ & $\times$ & $\times$  \\
1.5 & 0.35 & 0.005 & $\times$ & $\times$ & $\times$  \\
2.0 & 0.6 & 0.1 & $\times$ & $\times$ & $\times$  \\
2.5 & 0.7 & 0.25 & 0.03 & $\times$ & $\times$  \\
3.0 & 0.9 & 0.35 & 0.1 & 0.002 & $\times$  \\
3.5 & 1.0 & 0.5 & 0.2 & 0.04 & $\times$  \\
4.0 & 1.1 & 0.6 & 0.3 & 0.1  & 0.02  \\
4.5 & 1.25 & 0.7 & 0.4 & 0.2*  & 0.3*   \\
10.0 & 2.0 & 1.4 & 1.3*  & 1.3*  & 1.3*  \\
\hline

\end{tabular}
\caption{Integrated $e$-folding number in the shocked molecular cloud.
The cross mark indicates that the gas is fully stable. The asterisk indicates that the dust-gas collisional cooling
is the dominant cooling in the model.}
\label{tab:molecular-result}
\end{center}
\end{table}

\clearpage


\clearpage

\clearpage


\clearpage



\clearpage




\end{document}